\documentclass[a4paper, 10pt, conference]{ieeeconf}
\IEEEoverridecommandlockouts
\usepackage{amsfonts}
\usepackage{dsfont}
\usepackage{mathrsfs}
\usepackage{amssymb}
\usepackage{bbm}
\usepackage{epsfig}
\usepackage{amsmath}
\usepackage{float}
\usepackage{color}
\usepackage{cite}
\usepackage[english]{babel}

\newtheorem{defn}{Definition}

\newtheorem{lem}{Lemma}

\newtheorem{prop}{Proposition}

\title{\bf Randomized Consensus with Attractive and Repulsive Links}
\date{}
\author{Guodong Shi, Alexandre Proutiere, Mikael  Johansson, and Karl H. Johansson\thanks{The authors are with ACCESS Linnaeus Centre, School of Electrical Engineering,
Royal Institute of Technology, Stockholm 10044, Sweden. e-mail: guodongs@kth.se, alepro@kth.se, mikaelj@kth.se, kallej@kth.se. This work has been supported in part by the Knut
    and Alice Wallenberg Foundation, the Swedish Research Council and  KTH SRA TNG.}}

\begin{document}
 \maketitle

\begin{abstract}
We study convergence properties of a randomized consensus algorithm over a graph with both attractive and repulsive links. At each time instant, a node is randomly selected to interact with a random neighbor. Depending on if the link between the two nodes belongs to a given subgraph of attractive or repulsive links, the node update follows a standard attractive weighted average or a repulsive weighted average, respectively. The repulsive update has the opposite sign of the standard consensus update. In this way, it counteracts the consensus formation and can be seen as a model of link faults or malicious attacks in a communication network, or the impact of trust and antagonism in a social network. Various probabilistic convergence and divergence conditions are established. A threshold condition for the strength of the repulsive action is given for convergence in expectation: when the repulsive weight crosses this threshold value, the algorithm transits from convergence to divergence. An explicit value of the threshold is derived for classes of attractive and repulsive graphs. The results show that a single repulsive link can sometimes drastically change the behavior of the consensus algorithm. They also explicitly show how the robustness of the consensus algorithm depends on the size and other properties of the graphs.
\end{abstract}


\section{Introduction}

Distributed consensus algorithms have been serving as basic models of information dissemination and aggregation over complex networks throughout a wide area of sciences including social sciences, engineering, and biology,  e.g., opinion dynamics  over social networks \cite{social1, social2, daron,como,julien1}, parallel computation and data fusion for sensor networks \cite{cs2,cs3,tpds,soh},   formation control in robotic  networks \cite{tsi,jad03,saber,ren}, and  flocking of animal groups \cite{flock1,flock2}.

 In a typical consensus algorithm, a node collects information from a subset of nodes in the network called neighbors and updates its state following an ``attractive" rule, a convex combination of its own and the neighbors' previous states. The neighbor relations and communication are often random, which lead to random consensus algorithms. The convergence of random consensus algorithms have been extensively studied in the literature \cite{hatano}--\cite{bamieh}. A great advantage for distributed consensus seeking lies in the fact that it is robust with respect to link failures and communication noise \cite{litao,huang,kar2,shirobust1,shirobust2}. Moreover, due to the attractive update, different probabilistic convergence concepts often coincide for random consensus algorithms \cite{chen11}.

Few works have discussed the influence of ``repulsive" links in the network on the consensus formation despite the many motivations for doing so.
In social networks, signed graphs were introduced for formulating the tensions and conflicts between individuals. Links representing interpersonal connection were associated with a sign which indicates if the
mutual relationship is friendship
or hostility \cite{sign1,sign2,sign3}. In sensor networks, the communication links can be taken by attackers so that data can be injected to oppose consensus \cite{attack2}. In  collaborative networks, malicious users may exist whose
objective is to damage the network and increase the cost incurred
by the legitimate users \cite{baras}.

In \cite{alta}, a class of antagonistic interactions modeled as negative weights in the update law were studied in a continuous-time setting, and necessary and sufficient conditions were derived for consensus over the network in absolute value. In \cite{shi}, a randomized model was formulated where each node executes an attraction, repulsion or neglect update at random when meeting other nodes.

In this paper, we study a  random  consensus model  with both attractive and repulsive links in the underlying communication network. Contrary to the model in~\cite{shi}, where attractive and repulsive updates are selected at random, the model in this paper allows the update type to be selected based on predetermined inter-node relations. We use a gossiping model to define how nodes are selected for updating \cite{gossip1,gossip2,boyd,shicdc}.  In each time slot, a random node is selected to interact with a random neighbor. The node updates its state following standard attractive weighted average or repulsive weighted average,  determined by whether the link is attractive or repulsive. Our main contributions are the following.

\begin{itemize}
 \item We establish various conditions for convergence or divergence in expectation, in mean square, and almost surely. In contrast to the standard consensus model without repulsive updates, some fundamental differences show up in these probabilistic  modes.

     \item We show that under mild assumptions there is a threshold value for the strength of the repulsive action for which the convergence in expectation changes: when the repulsive weight crosses this threshold, the randomized consensus algorithm transits from convergence to divergence. The explicit value of the threshold is derived for classes of attractive and repulsive graphs.

         \item We establish a no-survivor theorem for almost sure divergence, which indicates that a single repulsive link can drastically change the behavior of the overall network.
\end{itemize}

The paper is organized as follows. Section II introduces the network model and defines the problem of interest. Section~III discusses convergence and divergence in expectation and shows that there is a threshold value for phase transition. Example graphs are studied and explicit threshold values  are derived. Sections IV and V present mean-square and almost sure convergence and divergence conditions, respectively. Finally concluding remarks are given in Section VI.
\section{Problem Definition}
In this section, we present the considered network model and define the problem of interest.
We first  recall  some  basic definitions from graph theory \cite{god} and stochastic matrices \cite{lat}. A directed graph (digraph) $\mathcal
{G}=(\mathcal {V}, \mathcal {E})$ consists of a finite set
$\mathcal{V}=\{1,\dots,n\}$ of nodes and an arc set
$\mathcal {E}\subseteq \mathcal{V}\times\mathcal{V}$.  An element $e=(i,j)\in\mathcal {E}$ is  an
{\it arc}  from node $i\in \mathcal{V}$  to $j\in\mathcal{V}$. A digraph $\mathcal {G}$ is  bidirectional if for every two nodes $i$ and $j$, $(i,j)\in \mathcal{E}$  if and only if $(j,i)\in \mathcal{E}$.
 A finite square matrix $M=[m_{ij}]\in\mathds{R}^{n\times n}$ is called {\em stochastic} if $m_{ij}\geq 0$ for all $i,j$ and $\sum_j m_{ij}=1$ for all $i$.  A stochastic matrix $M$ is  {\em doubly stochastic} if also $M^T$ is  stochastic. Let $P=[p_{ij}]\in\mathds{R}^{n\times n}$ be a matrix with nonnegative entries. We can associate a unique digraph  $\mathcal{G}_P=(\mathcal{V},\mathcal{E}_P)$ with $P$ on node set $\mathcal{V}$ such that $(j,i)\in\mathcal{E}_P$ if and only if $p_{ij}>0$. We call $\mathcal{G}_P$ the {\em induced graph} of $P$.

\subsection{Node Pair Selection}

Consider a network  with node set $\mathcal{V}=\{1,\dots,n\}$, $n\geq3$.  Let the digraph $\mathcal {G}_0=(\mathcal{V},\mathcal{E}_0)$ denote the {\it underlying}  graph of the considered network. The underlying graph indicates  potential interactions between nodes. We use the asynchronous time model introduced in \cite{boyd} to describe node  interactions. Each node meets other nodes at independent time instances  defined  by a  rate-one Poisson process. This is to say, the inter-meeting times at each node follows a rate-one  exponential distribution. Without loss of generality, we can assume that  at most one node is active at any given instance. Let $x_i(k)\in\mathds{R}$ denote the state (value) of node $i$ at the $k$'th meeting slot among all the nodes.

  Node interactions are characterized by an $n\times n$ matrix $P=[p_{ij}]$, where $p_{ij}\geq0$ for all $i,j=1,\dots,n$ and $p_{ij}>0$ if and only if $(j,i)\in \mathcal{E}_0$. We assume $P$ to be a stochastic matrix. Without loss of generality we suppose $p_{ii}=0$ for all $i$. In other words, the underlying graph $\mathcal{G}_0$  the induced graph of the  matrix $P$. The meeting process is defined as follows.

\begin{defn}[Node Pair Selection] Independent of time and node state, at time $k\geq0$,
\begin{itemize}
\item[(i)]  A node $i\in\mathcal{V}$ is drawn    with probability $1/n$;
  \item[(ii)] Node $i$ picks node $j$ with probability $p_{ij}$.
\end{itemize}
In this way, we say arc $(j,i)$ is selected.
\end{defn}

\subsection{Attractive and Repulsive Graphs}
 We assign a partition of the underlying graph  $\mathcal{G}_0$ into two disjoint subgraphs, $\mathcal{G}_{{\rm att}}$ and $\mathcal{G}_{\rm rep}$,  namely, the attractive graph and the repulsive graph. To be precise, $\mathcal{G}_{\rm att}=(\mathcal{V}, \mathcal{E}_{\rm att})$ and $\mathcal{G}_{\rm rep}=(\mathcal{V}, \mathcal{E}_{\rm rep})$ are two graphs over node set $\mathcal{V}$ satisfying $\mathcal{E}_{\rm att} \cap\mathcal{E}_{\rm rep} =\emptyset$ and $\mathcal{E}_{\rm att} \cup\mathcal{E}_{\rm rep} =\mathcal{E}_0$. Under this graph partition the node pair selection matrix  $P$ can be naturally written as $P=P_{\rm att}+P_{\rm rep}$, for which $\mathcal{G}_{\rm att}$ is the induced graph of $P_{\rm att}$, and $\mathcal{G}_{\rm rep}$ is the induced graph of $P_{\rm rep}$.

Suppose  arc  $(j,i)$ is selected  at time $k$. Node $j$ keeps its previous state, and node $i$ updates its state following the rule:
\begin{itemize}
\item[(i)] {\it (Attraction)} If $(j,i)\in\mathcal{E}_{\rm att} $, node $i$ updates as a weighted average with $j$:
\begin{align}\label{trust}
x_i(k+1)=(1- \alpha_k)x_i(k)+ \alpha_k x_j(k),
\end{align}
where $0\leq \alpha_k \leq 1$.

\item[(ii)] {\it (Repulsion)} If $(j,i)\in\mathcal{E}_{\rm rep} $, node $i$  updates as a weighted average with $j$, but with a negative coefficient:
\begin{align}
x_i(k+1)=(1+\beta_k)x_i(k)- \beta_kx_j(k),
\end{align}
where $\beta_k\geq 0$.
\end{itemize}

\subsection{Problem of Interest}


We introduce  the following definition.
\begin{defn} (i) Consensus convergence for initial value $x^0\in \mathds{R}^{n}$ is achieved
\begin{itemize}
\item in \emph{expectation}   if
$\lim_{k\rightarrow \infty} \big|\mathbf{E} \big[x_i(k)-x_j(k) \big]\big|=0$ for all $i$ and $j$;
\item  in \emph{mean square}    if $\lim_{k\rightarrow \infty} \mathbf{E} \big[x_i(k)-x_j(k) \big]^2=0$ for all $i$ and $j$;
\item \emph{almost surely}   if $
\mathbf{P}\big(\lim_{k\rightarrow \infty} |x_i(k)-x_j(k)|=0\big)=1
$
for all $i$ and $j$.
\end{itemize}
(ii) Consensus  divergence for initial value $x^0\in \mathds{R}^{n}$ is achieved \begin{itemize}
\item in \emph{expectation}   if
$\limsup_{k\rightarrow \infty}  \max_{i,j} \big|\mathbf{E} \big[x_i(k)-x_j(k) \big]\big|=\infty$;
\item in  \emph{mean square} if
$\limsup_{k\rightarrow \infty}  \max_{i,j} \mathbf{E} \big[x_i(k)-x_j(k) \big]^2=\infty$;
\item {\em almost surely} if for all $M\geq 0$,
$\mathbf{P}\big(\limsup_{k\rightarrow \infty} \max_{i,j} |x_i(k)-x_j(k)|>M\big)=1$.
\end{itemize}
\end{defn}

Global consensus convergence in expectation, in mean square, and almost surely are defined when the convergence holds for all $x^0$ in each of the three cases.

\section{Convergence vs. Divergence in Expectation}
The considered randomized  algorithm can be expressed as
 \begin{align}\label{1}
 x(k+1)=W(k)x(k),
 \end{align}
where $W(k)$ is the random matrix satisfying
\begin{align}\label{2}
\begin{cases}
\mathbf{P}\Big(W(k)= I-\alpha_k e_i(e_i-e_j)^T\Big)=\frac{p_{ij}}{n}, &  (j,i)\in \mathcal{E}_{\rm att}\\
\mathbf{P}\Big(W(k)= I+\beta_k e_i(e_i-e_j)^T\Big)=\frac{p_{ij}}{n}, &  (j,i)\in \mathcal{E}_{\rm rep}
\end{cases}
\end{align}
 with $e_m=(0 \dots 0\  1\  0 \dots 0)^T$ denoting the $n\times1$ unit vector whose $m$'th component is $1$.

Denote $D_{\rm att}={\rm diag}(d_1 \dots d_n)$ with $d_i= \sum_{j=1}^n [P_{\rm att}]_{ij}$. Denote also $D_{\rm rep}={\rm diag}(\bar{d}_1 \dots \bar{d}_n)$ with $\bar{d}_i= \sum_{j=1}^n [P_{\rm rep}]_{ij}$. Define $ L_{\rm att}= D_{\rm att}- P_{\rm att}$ and $ L_{\rm rep}= D_{\rm rep}- P_{\rm rep}$. Then $ L_{\rm att}$ and  $ L_{\rm rep}$ represent the (weighted) Laplacian matrices of the attractive graph $ \mathcal{G}_{\rm att}$ and repulsive graph  $ \mathcal{G}_{\rm rep}$, respective.
After some simple algebra it can be shown that
\begin{align}
\mathbf{E} W(k)= I- \frac{\alpha_k}{n} L_{\rm att}+ \frac{\beta_k}{n} L_{\rm rep}\doteq \bar{W}_k.
\end{align}

\subsection{General  Conditions}

Introduce $y_i(k)=x_i(k)-\frac{1}{n}\sum_{i=1}^n x_i(k)$. Then  $y(k)=(y_1(k)\dots y_n(k))^T=x(k)-\frac{\mathbf{1} \mathbf{1}^T}{n} x(k)$ with $\mathbf{1}=(1 \dots  1)^T$ denoting the $n\times1$ vector each component of which is $1$. Then it is straightforward to see that  consensus convergence in expectation is achieved  if and only if $\lim_{k\rightarrow \infty}\mathbf{E} y(k) =0$, and consensus divergence  in expectation is achieved  if and only if  $\limsup_{k\rightarrow \infty}\big|\mathbf{E} y(k)\big| =\infty$.

Let $\lambda_{\rm max}(A)$ denote the largest eigenvalue for a symmetric matrix $A$. We have the following result.

\begin{prop}\label{prop1}  Global consensus convergence in expectation is achieved if $\prod _{k=0}^\infty \lambda_{\rm max}\big( \bar{W}_k^T(I-\frac{\mathbf{1} \mathbf{1}^T}{n}) \bar{W}_k\big)=0$.
\end{prop}
{\it Proof.} Since $\mathbf{E} W(k)$ is a stochastic matrix and the node pair selection is independent of the node states, we obtain
\begin{align}\label{3}
&\mathbf{E} y(k+1) =(I-{\mathbf{1} \mathbf{1}^T}/{n}) \bar{W}_k \mathbf{E} x(k)\nonumber\\
&=(I-{\mathbf{1} \mathbf{1}^T}/{n}) \bar{W}_k \mathbf{E} y(k)+(I-{\mathbf{1} \mathbf{1}^T}/{n}) \bar{W}_k{\mathbf{1} \mathbf{1}^T}/{n}\mathbf{E} x(k)  \nonumber\\
&=(I-{\mathbf{1} \mathbf{1}^T}/{n}) \bar{W}_k \mathbf{E} y(k)+(I-{\mathbf{1} \mathbf{1}^T}/{n}) {\mathbf{1} \mathbf{1}^T}/{n}\mathbf{E} x(k)\nonumber\\
&=(I-{\mathbf{1} \mathbf{1}^T}/{n}) \bar{W}_k \mathbf{E} y(k).
\end{align}
Thus, noticing that $(I-\frac{\mathbf{1} \mathbf{1}^T}{n})^2=(I-\frac{\mathbf{1} \mathbf{1}^T}{n})$, we have
\begin{align}
\big|\mathbf{E} y(k+1)\big|
&=\big|(I-{\mathbf{1} \mathbf{1}^T}/{n}) \bar{W}_k \mathbf{E} y(k)\big| \nonumber\\
&\leq  \|(I-{\mathbf{1} \mathbf{1}^T}/{n}) \bar{W}_k\|_2  \big|\mathbf{E} y(k)\big| \nonumber\\
&= \sqrt{\lambda_{\max}\big( \bar{W}_k^T(I-\frac{\mathbf{1} \mathbf{1}^T}{n}) \bar{W}_k\big)} \big| \mathbf{E} y(k)\big|,
\end{align}
where $\|\cdot\|_2$ denotes the spectral norm. The desired conclusion follows.   \hfill$\square$


 When $P_{\rm att}$ and $P_{\rm rep}$ are symmetric, an upper bound for  $\sqrt{ \lambda_{\max}\big( \bar{W}_k^T(I-\frac{\mathbf{1} \mathbf{1}^T}{n}) \bar{W}_k\big)}$ can be easily computed with the help of Weyl's inequality. We propose  the following result.

\begin{prop}\label{coro1} Suppose  both $P_{\rm att}$ and $P_{\rm rep}$ are symmetric. Global consensus convergence in expectation is achieved if $$
\prod _{k=0}^\infty \Big(1-\frac{\alpha_k}{n}\lambda_2(L_{\rm att})+\frac{\beta_k}{n}\lambda_{\max}(L_{\rm rep})\Big)=0
$$
where $\lambda_2(L_{\rm att})$ is the second largest eigenvalue of $L_{\rm att}$.
\end{prop}
{\it Proof.}
We have \begin{align}
&\sqrt{ \lambda_{\max}\big( \bar{W}_k^T(I-\frac{\mathbf{1} \mathbf{1}^T}{n}) \bar{W}_k\big)}\nonumber\\
&=\lambda_{\max}\big(\bar{W}_k-\frac{\mathbf{1} \mathbf{1}^T}{n}\big)\nonumber\\
&\leq\lambda_{\max}\big( I-\frac{\mathbf{1} \mathbf{1}^T}{n}-\frac{\alpha_k}{n}L_{\rm att}\big)+\lambda_{\max}(L_{\rm rep})\nonumber\\
&=1-\frac{\alpha_k}{n}\lambda_2 (L_{\rm att})+\frac{\beta_k}{n}\lambda_{\rm max}(L_{\rm rep}),
\end{align}
where the inequality holds from Weyl's inequality.
The desired conclusion follows directly from Proposition \ref{prop1}. \hfill$\square$

When $\alpha_k$ and $\beta_k$ are  time invariant, i.e., there are two constants $0\leq \alpha\leq 1$, $\beta\geq 0$ such that $\alpha_k\equiv \alpha$ and  $\beta_k\equiv \beta$ for all $k$, based on (\ref{3}), the consensus convergence in expectation is equivalent with the stability of the following LTI system:
$$
\mathbf{E} y(k+1)= (I-{\mathbf{1} \mathbf{1}^T}/{n}) \bar{W}\mathbf{E} y(k)
$$
where $\bar{W}=I- \frac{\alpha}{n} L_{\rm att}+ \frac{\beta}{n} L_{\rm rep}$. Consequently, letting  $\rho(A)$ represent the spectral radius  for a matrix $A$, i.e., the largest eigenvalue in magnitude, we have the following result.
\begin{prop}\label{coro2}  Assume that there are two constants $0\leq \alpha\leq 1$, $\beta\geq 0$ such that $\alpha_k\equiv \alpha$ and  $\beta_k\equiv \beta$ for all $k$.

(i). Global consensus convergence in expectation is achieved if and only if $\rho\big((I-\frac{\mathbf{1} \mathbf{1}^T}{n}) \bar{W}\big)<1$,

(ii).  Consensus divergence in expectation is achieved for almost all initial values if and only if $\rho\big((I-\frac{\mathbf{1} \mathbf{1}^T}{n}) \bar{W}\big)>1$.
\end{prop}

\subsection{Phase Transition }
Define
$
f(\alpha,\beta)\triangleq \rho\big((I-\frac{\mathbf{1} \mathbf{1}^T}{n}) \bar{W}\big).
$
We present the following result.

\begin{prop}\label{prop2}  Suppose $\mathcal{G}_{\rm att}$ has a spanning tree and $\mathcal{G}_{\rm rep}$ contains at least one link. Also assume that either of the following two conditions holds:

(i) $L_{\rm att} L_{\rm rep}=L_{\rm rep}L_{\rm att}$;

(ii)  $P_{\rm att}$ and $P_{\rm rep}$ are symmetric.

  Then for any fixed $\alpha \in(0,1]$, there exists a threshold value $\beta_\star(\alpha) \geq 0$ such that
\begin{itemize}
\item Global consensus convergence in expectation, i.e., $f(\alpha,\beta)<1$,  is achieved if  $0\leq \beta<\beta_\star $;

\item Consensus divergence  in expectation  for almost all initial values,  i.e., $f(\alpha,\beta)>1$, is achieved  if  $ \beta>\beta_\star $.
\end{itemize}
\end{prop}


 When both $P_{\rm att}$ and $P_{\rm rep}$ are symmetric, it turns out that some monotonicity can be established for $f$.
\begin{prop} \label{prop6} Suppose both $P_{\rm att}$ and $P_{\rm rep}$ are symmetric.
Then $f(\alpha,\cdot)$ is non-increasing  in  $\alpha$ for $\alpha\in[0,1]$; $f(\cdot,\beta)$ is non-decreasing in  $\beta$ for $\beta\in[0,\infty)$.
\end{prop}


The proofs  of Propositions \ref{prop2} and \ref{prop6} can be found in appendix.

\subsection{Examples:  Threshold Value}

We first consider the case when the underlying graph $\mathcal{G}_0$ is the complete graph ${K}_n$ and each link is selected with equal probability at any time step. We have the following results.

\begin{prop}\label{prop3} Suppose $P=\frac{1}{n-1}({\mathbf{1} \mathbf{1}^T-I})$.  Let $(\mathcal{G}_{\rm att}, \mathcal{G}_{\rm rep})$ be a given bidirectional  attraction-repulsion  partition. Then we have
$$
\beta_\star= \max \Big\{ \big (\frac{n}{(n-1)\lambda_{\rm max}(L_{\rm rep})}-1\big)\alpha, 0\Big\}.
$$
\end{prop}

 The proof of Proposition \ref{prop3} can be obtained straightforwardly from the following key lemma which indicates that the Laplacian matrix of the complete graph ${K}_n$ commutes with that of any other bidirectional graph.

\begin{lem}\label{lem2}
Let  ${K}_n$ be the complete graph and ${G}$ be any bidirectional graph. Then there always holds
$
L_{K_n}L_G=L_GL_{K_n},$
where $L_{K_n}$ and $L_G$ are the Laplacian matrices of $K_n$ and $G$, respectively.
\end{lem}

When the repulsive graph $\mathcal{G}_{\rm rep}$ is formed by the undirected  Erd\"{o}s-R\'{e}nyi random graph $\mathcal{G}(n,p)$ in the sense that for every unordered pair $\{i,j\}$, $(i,j)$ and $(j,i)$ are repulsive links with probability $p$. This gives us a sequence of random variables
$$
\xi_n=\rho\big((I-\frac{\mathbf{1} \mathbf{1}^T}{n}) \bar{W}\big), \ \ n=1,2,3,\dots.
$$
Note that  induced by $\{\xi_n\}_1^\infty$, the consensus convergence or divergence forms a well-defined random sequence indexed by $n$. We propose the following result.
\begin{prop}\label{prop4}
  Suppose $P=\frac{1}{n-1}({\mathbf{1} \mathbf{1}^T-I})$. Fix $\alpha_k\equiv \alpha\in(0,1]$ and $\beta_k\equiv \beta\in(0,\infty)$. Let $\mathcal{G}_{\rm rep}$ be formed by the undirected  Erd\"{o}s-R\'{e}nyi random graph $\mathcal{G}(n,p)$.  Then
  $$
  p_\star=\frac{\alpha}{\alpha+\beta}
  $$
  is a threshold value regarding the consensus convergence or divergence. To be precise, we have,
  \begin{itemize}
  \item[a)] When $p<p_\star$, global consensus convergence in expectation is achieved in probability,  i.e., $\lim_{n\rightarrow \infty}\mathbf{P}(\xi_n <1)=1$;
    \item[b)] When $p>p_\star$, consensus divergence in expectation for almost all initial values is achieved in probability,  i.e., $\lim_{n\rightarrow \infty}\mathbf{P}(\xi_n >1)=1$.
  \end{itemize}
\end{prop}
 The result follows directly from the following lemma.
\begin{lem}\label{lem3}\cite{ding}
Let $\Delta_n$ be the Laplacian of the Erd\"{o}s-R\'{e}nyi random graph $\mathcal{G}(n,p)$. Then
$\frac{\lambda_{\rm max}(\Delta_n)}{pn}\rightarrow 1$ in probability.
\end{lem}


Next, we discuss the other extreme case when the underlying communication graph is the ring graph, $R_n$, which is nearly the most {\it sparse} connected graph. We present the following result.

\begin{prop} Denote $A_{R_n}$ as the adjacency matrix of $R_n$.  Suppose $P=A_{R_n}/2$.  Let $(\mathcal{G}_{\rm att}, \mathcal{G}_{\rm rep})$ be a given bidirectional  attraction-repulsion  partition with $ \mathcal{G}_{\rm rep}\neq \emptyset$.  Then $\beta_\star\leq \alpha$ for all $n$.
\end{prop}
{\it Proof.}
It is well known that  $L_{R_n}$ has eigenvalues $2-2\cos(2\pi k/n), 0\leq k\leq n/2$. On the other hand, we have $\lambda_{\rm max}(L_{\rm rep})=1$. Based on Weyl's inequality, we obtain
\begin{align}
&\rho\big((I-\frac{\mathbf{1} \mathbf{1}^T}{n}) \bar{W}\big)\nonumber\\
&\geq \lambda_{\rm min} \Big(I- \frac{\alpha}{2n} L_{R_n}-\frac{\mathbf{1} \mathbf{1}^T}{n} \Big) +\frac{\alpha+\beta}{n}\lambda_{\rm max}(L_{\rm rep})\nonumber\\
&= 1-\frac{\alpha(1- \cos(2 \pi \lfloor \frac{n}{2}\rfloor / n) }{n}+\frac{\alpha +\beta }{n}\nonumber\\
&\geq 1+ \frac{\beta-\alpha}{n}.
\end{align}
This means that $\rho\big((I-\frac{\mathbf{1} \mathbf{1}^T}{n}) \bar{W}\big)>1$ whenever $\beta>\alpha$, which proves the desired conclusion. \hfill$\square$

\section{Convergence vs. Divergence in Mean Square}
This subsection discusses the mean square convergence and divergence for the considered algorithm.  With Cauchy-Schwarz inequality,  it holds that
\begin{align}\label{14}
|y_i|^2= \frac{1}{n^2}\Big|\sum_{i=1}^n(x_i-x_j)\Big|^2\leq \frac{1}{n} \sum_{i=1}^n \big|x_i-x_j \big|^2.
\end{align}
Moreover, we also have
\begin{align}\label{15}
\big|x_i-x_j\big|^2=\big|y_i-y_j\big|^2\leq 2\big(|y_i|^2+|y_j|^2\big).
\end{align}
Therefore,  consensus convergence in mean square  is achieved  if and only if $\lim_{k\rightarrow \infty}\mathbf{E}| y(k)|^2 =0$, and  consensus divergence  in mean square  is achieved  if and only if  $\limsup_{k\rightarrow \infty}\mathbf{E} |y(k)|^2 =\infty$.

We present the following result.

\begin{prop}  (i) Global consensus convergence in mean square is achieved if $\prod _{k=0}^\infty \lambda_{\rm max}\big( \mathbf{E} \big[{W}_k^T(I-\frac{\mathbf{1} \mathbf{1}^T}{n}) {W}_k \big]\big)=0$.

 (ii) Consensus divergence  in mean square is achieved for almost all initial values if
 $
 \prod _{k=0}^\infty  \lambda_2 \big( \mathbf{E} \big[{W}_k^T(I-\frac{\mathbf{1} \mathbf{1}^T}{n}) {W}_k \big]\big)=\infty$, where  $\lambda_2$ is the second largest eigenvalue.
\end{prop}
{\it Proof.} Noticing that $W_k$ is a stochastic matrix for all possible samples, we obtain
\begin{align}
&\mathbf{E} \Big(\big| y(k+1)\big|^2 \Big |y(k)\Big) \nonumber\\
&=\mathbf{E}\Big( y(k)^T W_k^T (I-\frac{\mathbf{1} \mathbf{1}^T}{n}) W_k y(k)  \Big |y(k) \Big)\nonumber\\
&=\mathbf{E}\Big( y(k)^T \mathbf{E}\big(W_k^T (I-\frac{\mathbf{1} \mathbf{1}^T}{n}) W_k\big) y(k)  \Big |y(k)\Big)\nonumber\\
&\leq \lambda_{\rm max}\big( \mathbf{E} \big[{W}_k^T(I-\frac{\mathbf{1} \mathbf{1}^T}{n}) {W}_k \big]\big) | y(k)|^2,
\end{align}
where the second equality holds from the fact that $W_k$ is independent of time and the node states, and the inequality holds from the Rayleigh-Ritz theorem. Similarly we have
\begin{align}
\mathbf{E}\Big(\big| y(k+1)\big|^2 \Big |y(k)\Big) &\geq \lambda_2\big( \mathbf{E} \big[{W}_k^T(I-\frac{\mathbf{1} \mathbf{1}^T}{n}) {W}_k \big]\big) | y(k)|_\mathrm{C}^2\nonumber\\
&=\lambda_2\big( \mathbf{E} \big[{W}_k^T(I-\frac{\mathbf{1} \mathbf{1}^T}{n}) {W}_k \big]\big)  | y(k)|^2, \nonumber
\end{align}
where $\mathrm{C}\doteq \{y:y_1=\dots=y_n\}$, and the equality holds from the fact that $\mathbf{1}^Ty(k)=0$ for all $k$.
The desired conclusion follows immediately.  \hfill$\square$


\section{Almost Sure Convergence vs. Divergence}
We move to the discussion on almost sure consensus convergence and divergence in this subsection.
First we study a special case when $\alpha_k\equiv 1$. The following result holds.
\begin{prop}
Suppose $\alpha_k\equiv 1$ and $\mathcal{G}_{\rm att}$ has a spanning tree. Then for any sequence of $\{\beta_k\}_0^\infty$, global consensus is achieved almost surely in finite time, i.e., $$
\mathbf{P}\Big(\exists K, {\rm s.t.},  x_i(k)=x_j(k), i,j\in\mathcal{V}, k\geq K\Big)=1.
$$
Denoting $\mathrm{T}_0=\inf_{k}\big \{  x_i(k)=x_j(k),i,j\in\mathcal{V}\big\}$ as the initial time when consensus is reached, we have $\mathbf{E} \mathrm{T}_0 \leq (n-1)\big(\frac{n}{p_\ast}\big)^{n-1}$, where $p_\ast=\min\{p_{ij}: p_{ij}>0\}$.
\end{prop}
{\it Proof.} Introduce
$$
m(k)=\min_{i\in \mathcal{V}} x_i(k); \quad M(k)=\max_{i\in \mathcal{V}} x_i(k).
$$
We define $\mathcal{M}(k)=M(k)-m(k)$. Following the considered algorithm $\mathcal{M}(k)$ is a Markov chain with nonnegative states. The structure of the randomized algorithm gives
$$
\mathbf{P} \Big( \mathcal{M}(s)=0, s\geq k  \Big| \mathcal{M}(k)=0\Big)=1.
$$
Thus, zero is an absorbing state for $\mathcal{M}(k)$.

Since $\mathcal{G}_{\rm att}$ has a spanning tree, we can select a node $i_0$ which is a root node in  $\mathcal{G}_{\rm att}$. With  $\alpha_k\equiv 1$,  we have
\begin{align}
\mathbf{P}\Big(x_i(k+n-1)=x_{i_0}(k),\ i\in\mathcal{V}\Big)\geq \Big(\frac{p_\ast}{n}\Big)^{n-1},
\end{align}
which implies
\begin{align}\label{16}
\mathbf{P} \Big( \mathcal{M}(k+n-1)=0 \Big| \mathcal{M}(k)>0\Big)\geq \Big(\frac{p_\ast}{n}\Big)^{n-1}.
\end{align}
The Borel-Cantelli Lemma ensures that
$$
\mathbf{P} \Big( \exists k, \mathcal{M}\big(k(n-1)\big)=0 \Big| \mathcal{M}(0)>0\Big)=1,
$$
which proves the almost sure finite-time consensus.

With (\ref{16}), the upper bound  $(n-1)\big(\frac{n}{p_\ast}\big)^{n-1}$ of $\mathbf{E} \mathrm{T}_0 $ can be obtained by direct calculation of the expected value of the initial success time for a sequence of i.i.d. Bernoulli trials with success probability $\big(\frac{p_\ast}{n}\big)^{n-1}$. The proof is finished. \hfill$\square$

Clearly if $\{\beta_k\}_0^\infty$ is sufficiently large, both consensus divergence in expectation and in the mean square are possible when $\alpha_k\equiv 1$. Hence with repulsive links, the various notions of convergence are not equivalent, which contrasts with the case where all links are attractive.

\begin{prop}
Suppose $\mathcal{G}_{\rm att}$ has a spanning tree.  Global consensus convergence is achieved almost surely if

(i) there exists $\beta^\ast>0$ such that  $\beta_k\leq \beta^\ast$ for all $k$,

(ii) $0\leq \Phi_{k(n-1)}\leq 1$ with $\sum_{k=0}^\infty \Phi_{k(n-1)}=\infty$, where
$
\Phi_{s}=1- \big(1- \frac{ \prod_{k=s}^{s+n-1}\alpha_{k} }{2}\big)\big(\frac{p_\ast}{n}\big)^{n-1}- \big(1-\Big(\frac{p_\ast}{n}\big)^{n-1}\Big)\prod_{k=s}^{s+n-1} \big(1+\beta_k\big).$
\end{prop}
{\it Proof.} The proof is based on a similar martingale argument as \cite{aysal}. Let $i_0$ be a root node in  $\mathcal{G}_{\rm att}$. Take $k_0\geq 0$. Assume that $x_{i_0}(k_0)\leq \frac{1}{2}m(k_0)+\frac{1}{2}M(k_0)$. Since $i_0$ is a root node, there is node $i_1$ different from $i_0$ such that  $(i_0,i_1)\in\mathcal{E}_{\rm att}$. If arc $(i_0,i_1)$ is selected at time $k_0$, we have
\begin{align}
x_{i_1}(k_0+1)
&\leq (1-\alpha_{k_0})M(k_0)+\alpha_{k_0}\big(\frac{1}{2}m(k_0)+\frac{1}{2}M(k_0)\big)\nonumber\\
&=\frac{\alpha_{k_0}}{2}m(k_0)+(1-\frac{\alpha_{k_0}}{2})M(k_0).
\end{align}
Similarly, there is node $i_2$,  different from $i_0$ and $i_1$,  such that  at least one of $(i_0,i_2)\in\mathcal{E}_{\rm att}$ and  $(i_1,i_2)\in\mathcal{E}_{\rm att}$ holds. If $(i_0,i_1)$ is selected at time $k_0$, and either $(i_0,i_2)$ or  $(i_1,i_2)\in\mathcal{E}_{\rm att}$ is selected at time $k_0+1$,  we have
\begin{align}
&x_{i_2}(k_0+2)\leq (1-\alpha_{k_0+1})x_{i_2}(k_0+1)\nonumber\\
&+\alpha_{k_0+1}\max\{ x_{i_0}(k_0+1), x_{i_1}(k_0+1)\}\nonumber\\
&=\frac{\alpha_{k_0} \alpha_{k_0+1}}{2}m(k_0)+\big(1- \frac{\alpha_{k_0} \alpha_{k_0+1}}{2}\big) M(k_0).
\end{align}
The process can be continued since $\mathcal{G}_{\rm att}$ has a spanning tree, and with a proper choice of arcs in  $\mathcal{E}_{\rm att}$ for $k_0+2,\dots,k_0+n-2$, we have $m(k_0+n-1)=m(k_0)$ and
\begin{align}
x_{i}(k_0+n-1)&\leq \frac{ \prod_{k=k_0}^{k_0+n-1}\alpha_{k} }{2}m(k_0)\nonumber\\
&+\big(1- \frac{ \prod_{k=k_0}^{k_0+n-1}\alpha_{k} }{2}\big) M(k_0), \ i\in\mathcal{V},
\end{align}
which yield
\begin{align}\label{20}
\mathbf{P}\Big( \mathcal{M}(k_0+n-1)&\leq \big(1- \frac{ \prod_{k=k_0}^{k_0+n-1}\alpha_{k} }{2}\big) \mathcal{M} (k_0) \Big)\nonumber\\
& \geq \Big(\frac{p_\ast}{n}\Big)^{n-1}.
\end{align}
For the other case with $x_{i_0}(k_0)> \frac{1}{2}m(k_0)+\frac{1}{2}M(k_0)$, we can apply the same analysis on $z_i(k)$ with $z_i(k)=-x_i(k)$ and (\ref{18}) still holds.

On the other hand, the structure of the algorithm  ensures that
\begin{align}\label{19}
\mathbf{P}\Big( \mathcal{M}(k_0+n-1)&\leq \prod_{k=k_0}^{k_0+n-1} \big(1+\beta_k\big) \mathcal{M} (k_0) \Big)=1.
\end{align}
In light of (\ref{20}) and (\ref{19}), we obtain
\begin{align}\label{18}
\mathbf{E}\Big( \mathcal{M}(k_0+n-1)\Big| \mathcal{M} (k_0) \Big) \leq  \big[1-\Phi_{k_0}\big] \mathcal{M} (k_0).
\end{align}

We  invoke  the supermartingale convergence theorem to complete the final piece of the proof.
\begin{lem}\label{lem7}\cite{polyak} Let $V_k, k\geq 0$ be a sequence of nonnegative random variables with $\mathbf{E} V_0 <\infty$.  If
$$
\mathbf{E}\Big( V_{k+1}\Big| V_0,\dots,V_k\Big) \leq  (1-c_k)V_k
$$
with $c_k \in[0,1]$ and $\sum_{k=0}^\infty c_k= \infty$, then
$lim_{k\rightarrow \infty}V_k=0$ almost surely.
\end{lem}

With Lemma \ref{lem7} and (\ref{18}), we have $\lim_{k\rightarrow \infty} \mathcal{M}\big((n-1)k\big)=0$ almost surely if $0\leq \Phi_{k(n-1)}\leq 1$ and $\sum_{k=0}^\infty \Phi_{k(n-1)}=\infty$. Noticing the boundedness of $\beta_k$, the desired conclusion follows immediately.  \hfill$\square$

For almost sure divergence, we first present the following result which indicates that as long as almost sure divergence is achieved, then no node can ``survive" if the attractive graph is strongly connected.
\begin{prop}\label{prop7}
Suppose $\mathcal{G}_{\rm att}$ is strongly connected and consensus divergence is achieved almost surely. Suppose also there exists $\alpha_\ast>0$ such that $\alpha_k\geq \alpha_\ast$ for all $k$. Then
\begin{equation}
\mathbf{P}\Big(\limsup_{k\rightarrow \infty}  |x_i(k)-x_j(k)|>M_\ast\Big)=1
\end{equation}
for all $i$, $j$, and $M_\ast\geq 0$.
\end{prop}
{\it Proof.} Suppose consensus divergence is achieved almost surely. Fix $M_\ast>0$. Then we can find two nodes $i_\ast$ and $j_\ast$, such that almost surely, there exist a sequence $k_1<k_2<\dots$ satisfying $x_{i_\ast}(k_m)=m(k_m)$, $x_{j_\ast}(k_m)=M(k_m)$, and $ |x_{i_\ast}(k_m)-x_{j_\ast}(k_m)|=\mathcal{M}(k_m)\geq M_\ast$
for all $m=1,2,\dots$.

Now take two nodes  $i^\ast$ and $j^\ast$. Since $\mathcal{G}_{\rm att}$ is strongly connected, there are directed paths $i_\ast i_1\dots i_{\tau_1}i^\ast$ and $j_\ast j_1\dots j_{\tau_2}j^\ast$ in $\mathcal{G}_{\rm att}$, where $ 0\leq \tau_1,\tau_2\leq n-2$.  We impose a recursive argument to establish an upper bound for $x_{i^\ast}(k_m+(\tau_1+1)d_0)$. Take $\mu_0\in(0,1)$ and define $d_0=\inf\{d: (1-\alpha_{\ast})^{d}\leq \mu_0\}$.




 Suppose $(i_\ast, i_1)$ is selected at time steps $k_m,\dots,k_m+d_0-1$, $(i_1, i_2)$ is selected at time steps $k_m+d_0,\dots,k_m+2d_0-1$, etc.  We can obtain by recursive calculation that
\begin{align}\label{21}
&x_{i^\ast}\big(k_m+(\tau_1+1)d_0\big)
\leq \big(1-\mu_0\big)^{\tau_1+1} m(k_m)\nonumber\\
&+\Big(1-\big(1-\mu_0\big)^{\tau_1+1}\Big)M(k_m).
\end{align}

A lower  bound for $x_{j^\ast}\big(k_m+(\tau_1+\tau_2+2)d_0\big)$ can be established using the same argument as
\begin{align}\label{22}
&x_{j^\ast}\big(k_m+(\tau_1+\tau_2+2)d_0\big)
\geq \big(1-\mu_0\big)^{\tau_2+1} M(k_m)\nonumber\\
&+\Big(1-\big(1-\mu_0\big)^{\tau_2+1}\Big)m(k_m)
\end{align}
for a proper selection of arcs during time steps $k_m+(\tau_1+1)d_0,\dots,k_m+(\tau_1+\tau_2+2)d_0-1$.

We can compute the probability of such selection of sequence of arcs in the previous recursive estimate, and we  conclude from (\ref{21}) and (\ref{22}) that
\begin{align}\label{24}
&\mathbf{P}\Big( x_{j^\ast}\big(k_m+D_0\big)-x_{i^\ast}\big(k_m+D_0\big) \geq \big[ \big(1-\mu_0\big)^{\tau_2+1}\nonumber\\
&+\big(1-\mu_0\big)^{\tau_1+1}-1\big]\mathcal{M}(k_m)\geq\big[ \big(1-\mu_0\big)^{\tau_2+1}\nonumber\\
&+\big(1-\mu_0\big)^{\tau_1+1}-1\big]M_\ast\Big)\geq \Big(\frac{p_\ast}{n}\Big)^{(\tau_1+\tau_2+2)d_0}
\end{align}
where $D_0=(\tau_1+\tau_2+2)d_0$. Since $\mu_0$ is arbitrarily chosen we can always assume $\big(1-\mu_0\big)^{\tau_2+1}+\big(1-\mu_0\big)^{\tau_1+1}-1>1/2$. Thus, (\ref{24}) is reduced to
\begin{align}
&\mathbf{P}\Big( x_{j^\ast}\big(k_m+D_0\big)-x_{i^\ast}\big(k_m+D_0\big) \nonumber\\
&\geq \frac{1}{2}M_\ast\Big)\geq \Big(\frac{p_\ast}{n}\Big)^{(\tau_1+\tau_2+2)d_0},\ \ m=1,2,\dots.
\end{align}

Noting the fact that the events  $\big\{ x_{j^\ast}\big(k_m+D_0\big)-x_{i^\ast}\big(k_m+D_0\big) \geq \frac{1}{2}M_\ast\big\}$ are determined by the
node pair selection process, which is independent of time and node states, the Borel-Cantelli Lemma ensures that almost surely, we can select an infinite subsequence $k_{m_s}, s=1,2,\dots$, such that
$$
x_{j^\ast}\big(k_{m_s}+D_0\big)-x_{i^\ast}\big(k_{m_s}+D_0\big) \geq \frac{1}{2}M_\ast, \ \ 1,2,\dots.
$$
This has proved that
\begin{equation}
\mathbf{P}\Big(\limsup_{k\rightarrow \infty}  |x_{i^\ast}(k)-x_{j^\ast}(k)|>\frac{1}{2}M_\ast\Big)=1\  {\rm for\ all}\ M_\ast\geq 0.
\end{equation}
Since $i^\ast$ and $j^\ast$ are arbitrarily chosen, we have completed the proof. \hfill$\square$

Proposition \ref{prop7} shows that divergence is also propagated among the network between any two nodes.  Denoting $p^\ast=\max\{p_{ij}: p_{ij}>0\}$ and $E_0=|\mathcal{E}_{\rm att}|$. We end the discussion of this section by presenting the following almost sure divergence result.
\begin{prop}
Suppose $\mathcal{G}_{\rm rep}$ is weakly connected.  Global consensus convergence is achieved almost surely if

(i) there exists $\alpha^\ast<1$  such that $\alpha_k\leq\alpha^\ast$ for all $k$;

(ii) there exists $\beta_\ast>0$ such that $\beta_k\leq \beta_\ast$ for all $k$;

(iii)  there exists an integer $Z\geq 1$ such that $\sum_{m=0}^t Q(t)=O(t)$, where for $m=0,1,\dots,$
$
Q(m)\doteq\Big(\frac{p_\ast}{n}\Big)^{Z}\Big[\log \frac{1}{n-1}+\sum_{k=mZ}^{(m+1)Z-1}\log(1+\beta_{k})\Big]+ \Big(1- \big(1-\frac{p^\ast}{n})^{E_0Z}\Big) \Big[ \sum_{k=mZ}^{(m+1)Z-1} \log \big( 1-\alpha_k\big)\Big].$
\end{prop}
{\it Proof.} Since $\alpha_k<1$ for all $k$, the structure of the algorithm automatically implies that $\mathcal{M}(k)>0$ is a sure event for all $k$. As a result, we can well define a sequence of random variables $\zeta_k= \frac{\mathcal{M}(k+1)}{\mathcal{M}(k)},\ \  k=0,1,\dots$
as long as $\mathcal{M}(0)>0$.  From the considered randomized algorithm, it is easy to see that
\begin{align}\label{30}
\mathbf{P}\Big(\zeta_k=\frac{\mathcal{M}(k+1)}{\mathcal{M}(k)}\geq1-\alpha_k\Big)=1
\end{align}
and $\mathbf{P}\big(\zeta_k<1\big) \leq \mathbf{P}\big(\mbox{one attractive arcs is selected}\big)$.
Moreover, since $\mathcal{G}_{\rm rep}$ is weakly connected, for any time $k$, there must be  two nodes $i$ and $j$ with $(i,j)\in\mathcal{E}_{\rm rep}$ such that
$
|x_i(k)-x_j(k)|\geq \frac{\mathcal{M}(k)}{n-1}.$
Thus, if such $(i,j)$ is selected for time $k,k+1,k+Z-1$, we have
\begin{align}\label{31}
\mathcal{M}(k+Z)&\geq \big|x_i(k+Z)-x_j(k+Z)\big|\nonumber\\
&\geq  \frac{\mathcal{M}(k)}{n-1}\prod_{s=0}^{Z-1}(1+\beta_{k+s}).
\end{align}

Now we define a new sequence of random variables associated with the node pair selection process, $\chi_m^Z$, $m=0,1,\dots$, by
\begin{itemize}
\item[a)] $\chi_m^Z=\prod_{k=mZ}^{(m+1)Z-1} \big( 1-\alpha_k\big)$, if at least  one attractive arcs is selected for $k\in[mZ,\dots,(m+1)Z-1]$;
\item[b)] $ \chi_m^Z=\frac{1}{n-1}\prod_{k=mZ}^{(m+1)Z-1}(1+\beta_{k})$, if one repulsive arc $(i,j)$ satisfying $|x_i(mZ)-x_j(mZ)|\geq \frac{\mathcal{M}(mZ)}{n-1}$ is selcted for all $k\in[mZ,\dots,(m+1)Z-1]$
    \item[c)]   $ \chi_m^Z=1$ otherwise.
\end{itemize}
In light of (\ref{30}) and (\ref{31}), we  have
\begin{align}\label{36}
\mathbf{P}\Big(\prod_{k=mZ}^{(m+1)Z-1} \zeta_{k} \geq  \chi_m^Z, \ \ m=0,1,2,\dots\Big)=1.
\end{align}

From direct calculation according to the definition of $\chi_m^Z$, we have $\mathbf{E}\Big(\log \chi_m^Z\Big)\geq  Q(m)$.

We next invoke an argument from the strong law of large numbers to show that
\begin{align}\label{37}
\mathbf{P}\big(\limsup_{t\rightarrow \infty} \sum_{m=0}^{t}\log \chi_m^Z=\infty\big)=1.
\end{align}
Suppose there exist two constants $M_0\geq 0$ and $0<q<1$ such that
$\mathbf{P}\Big(\limsup_{t\rightarrow \infty} \sum_{m=0}^{t}\log \chi_m^Z\leq M_0\Big)\geq q$.
This leads to
\begin{align}\label{41}
\mathbf{P}\Big( \limsup_{t\rightarrow \infty} \frac{\sum_{m=0}^{t}\log \chi_m^Z }{t}\leq 0\Big)\geq q.
\end{align}

On the other hand, noting that the node updates are independent of time and node states, and that $\mathbf{V} (\log \chi_m^Z)$ is bounded in light of the bounds of $\alpha_k$ and $\beta_k$,  the strong law of large numbers  suggests that
\begin{align}
&\mathbf{P}\Big( \lim_{t \rightarrow \infty} \frac{1}{t}\sum_{m=0}^t\big(\log \chi_m^Z -Q(m)\big) \geq 0\Big)\nonumber\\
&\geq \mathbf{P}\Big( \lim_{m\rightarrow \infty} \frac{1}{t}\sum_{m=0}^t\big(\log \chi_m^Z-\mathbf{E}\log \chi_m^Z\big) = 0\Big)=1,\nonumber
\end{align}
which contradicts (\ref{41}) if $\limsup_{m\rightarrow \infty} \sum_{m=0}^t Q(m)=O(t) $.  Thus, (\ref{37}) is proved.

The final piece of the proof is based on (\ref{36}). With the definition of $\zeta_k$, (\ref{36}) yields
\begin{align}
&\mathbf{P}\Big( \log \mathcal{M}\big((t+1)Z\big) -\log\mathcal{M}\big(0\big)\nonumber\\
&= \sum_{k=0}^{(t+1)Z-1} \log \zeta_{k} \geq  \sum_{m=0}^t \log \chi_m^Z, \ \ t=0,1,2,\dots\Big)=1.
\end{align}
This gives us $\mathbf{P}\Big( \limsup_{t\rightarrow \infty} \log \mathcal{M}\big((t+1)Z\big) =\infty\Big)=1$
in light of (\ref{37}). The desired conclusion holds. \hfill$\square$

\section{Conclusions}
A randomized consensus algorithm  with both attractive and repulsive links has been studied under an asymmetric gossiping model.  The repulsive update was defined in the sense that a negative instead of a positive weight is imposed in the update. This model can represent the influence of certain link faults or malicious attacks in a communication network, or the spreading of trust and antagonism in a social network. We established various conditions for probabilistic convergence or divergence, and proved the existence of a phase-transition threshold  for convergence in expectation. An explicit value of the threshold was derived for classes of attractive and repulsive graphs. Future work includes the analysis for the symmetric update model and the structure optimization of the repulsive graph so that the maximum damage can be created for the network.

\section*{Appendix}
\begin{lem}\label{lem5}
Let $A$, $B$ be two matrices in $\mathds{R}^{n\times n}$.  Suppose $\|\cdot\|_\ast$ is any matrix norm. Then $g(\lambda)\triangleq \|A+\lambda B\|_\ast, \lambda\in \mathds{R}$ is a convex function in $\lambda$.
\end{lem}
{\it Proof.}  Noting that
\begin{align}
&g\big( t \lambda_1 +(1-t)\lambda_2\big)\nonumber\\
&=\Big\| t \big( A+ \lambda_1 B\big)+(1-t)\big(A + \lambda_2 B\big)\Big\|_\ast\nonumber\\
&\leq \Big\| t \big( A+ \lambda_1 B\big)\Big\|_\ast+\Big\|(1-t)\big(A + \lambda_2 B\big)\Big\|_\ast\nonumber\\
&=t g(\lambda_1)+(1-t) g(\lambda_2),
\end{align}
for all $t\in[0,1]$ and $\lambda_1,\lambda_2 \in\mathds{R}$, the lemma is proved.  \hfill$\square$

\begin{lem} \label{lem4} Let $K> 0$ be a positive constant. Let $\mathcal{M}_K$ be a subset of matrices in $\mathds{R}^{n\times n}$ such that

(i) $ |M_{ij}|\leq K$ for all $M\in \mathcal{M}_K$;

(ii) $M_1M_2=M_2M_1$ for all for all $M_1,M_2\in \mathcal{M}_K$.

Then for any $\epsilon>0$, there is a matrix norm $\|\cdot \|_\star$  such that
\begin{align}
\rho(A) \leq  \|A \|_\star \leq \rho(A) +\epsilon
\end{align}
for all $A\in \mathcal{M}_K$.
\end{lem}
{\it Proof.} Based on the simultaneous triangularization theorem (Theorem 2.3.3, \cite{horn}), there is a unitary matrix $U \in \mathds{C}^{n\times n}$ such that $U^\ast MU$ is upper triangular for every $M\in \mathcal{M}_K$ since $\mathcal{M}_K$ is a commuting family.   Similar to the proof of  Lemma 5.6.10 in \cite{horn}, we set $D_t={\rm diag}(t,t^2,\dots,t^n)$ and the desired matrix norm $\|\cdot\|_\star$ is obtained by
$
\|B\|_\star=\|D_tU^\ast BUD_t^{-1}\|_1$
taking $t$ sufficiently large. \hfill$\square$


\noindent {\it Proof of Proposition \ref{prop2}.} Case (i). Given $\epsilon>0$ and $\alpha\in[0,1]$. Take  $K_\ast> 0$. If $L_{\rm att} L_{\rm rep}=L_{\rm rep}L_{\rm att} $, it is easy to see that
$
\mathrm{M}_{K_\ast}\doteq \{(I-\frac{\mathbf{1} \mathbf{1}^T}{n}) \bar{W}:\beta\leq K_\ast\}
$
is a commuting family  satisfying the conditions in Lemma \ref{lem4}. Let $\|\cdot\|_\star$ be the matrix norm established in Lemma \ref{lem4}. We introduce
$g_\star(\beta)\triangleq \Big\|(I-\frac{\mathbf{1} \mathbf{1}^T}{n}) \bar{W}\Big\|_\star$.
When $\mathcal{G}_{\rm att}$ has a spanning tree, it is well known that $f(\alpha,0)= \rho\Big( (I-\frac{\mathbf{1} \mathbf{1}^T}{n})( I- \frac{\alpha}{n} L_{\rm att})  \Big) = 1- \delta< 1$
for some $0\leq \delta<1$. This gives us
\begin{align}\label{8}
g_\star(0)\leq  \rho\Big( (I-\frac{\mathbf{1} \mathbf{1}^T}{n})( I- \frac{\alpha}{n} L_{\rm att})  \Big)+\epsilon \leq 1- \delta+\epsilon
\end{align}
in light of  Lemma \ref{lem4}.

Now we take  $K_\ast\geq \beta_2>\beta_1 \geq 0$. We make the following claim.

{\it Claim.}  $f(\alpha,\beta_2)>1$ if $f(\alpha,\beta_1)>1$.

\vspace{1mm}
The fact that $f(\alpha,\beta_1)>1$ leads to that $g_\star(\beta_1)\geq f(\alpha,\beta_1)>1$.  According to Lemma \ref{lem5}, $g_\star(\cdot)$ is a convex function, which implies that
\begin{align}\label{9}
g_\star(\beta_1) \leq (1- \frac{\beta_1}{\beta_2})g_\star(0)+\frac{\beta_1}{\beta_2} g_\star(\beta_2).
\end{align}
With (\ref{8}) and (\ref{9}), we conclude that
\begin{align}
g_\star(\beta_2)\geq  1+ \Big(\frac{\beta_2}{\beta_1}-1\Big) \delta - \Big(\frac{\beta_2}{\beta_1}-1\Big) \epsilon,
\end{align}
which in turn yields that
\begin{align}\label{10}
 f(\alpha,\beta_2) \geq g_\star(\beta_2)- \epsilon\geq  1+ \Big(\frac{\beta_2}{\beta_1}-1\Big) \delta - \Big(\frac{\beta_2}{\beta_1}\Big) \epsilon,
\end{align}
again based on Lemma \ref{lem4}. Noting the fact that $\epsilon$ in (\ref{10}) can be chosen arbitrarily small,  and  that $\delta$, $\beta_2/\beta_1$, and $ f(\alpha,\beta)$ do not rely on the choice of $\|\cdot\|_\star$, we have proved that $ f(\alpha,\beta_2) >1$. The claim holds.

Next, we introduce
$
Z_\beta =\{\beta\geq 0:f(\alpha,\beta)>  1\big \}.$
First of all $Z_\beta$ is nonempty when  $L_{\rm rep}$ contains at least one link due to the simple fact that $
\lim_{\beta\rightarrow \infty}f(\alpha,\beta) =\infty.$ Secondly
\begin{align}\label{11}
\beta_\star\triangleq \inf  \big \{\beta\geq 0:f(\alpha,\beta)>  1\big \}
\end{align}
is a finite number in light of the claim we just established. It is straightforward to verify that $f(\alpha,\beta)>  1$  when $\beta >\beta_\star$. The fact that
$f(\alpha,\beta)< 1$  when $\beta <\beta_\star$ can be proved via a symmetric argument. This completes the proof for case (i).

Case (ii). From Lemma \ref{lem5} and the fact that $f(\alpha,\beta)=\|(I-\frac{\mathbf{1} \mathbf{1}^T}{n})\bar{W}\|_2$, $f(\alpha,\cdot)$ is a convex function in $\alpha$; $f(\cdot,\beta)$ is a convex function in $\beta$. The desired conclusion follows immediately. \hfill$\square$

\noindent {\it Proof of Proposition \ref{prop6}.} According to the proof of Proposition  \ref{coro1},  the eigenvalues of $(I-\frac{\mathbf{1} \mathbf{1}^T}{n}) \bar{W}$ are all nonnegative when $\alpha\in[0,1]$, $\beta\in[0,\infty)$, and both $P_{\rm att}$ and $P_{\rm rep}$. 
The Courant-Fischer Theorem  guarantees that
\begin{align}\label{5}
&\rho\big((I-\frac{\mathbf{1} \mathbf{1}^T}{n}) \bar{W}\big)=\max_{|z|=1} z^T \Big( I- \frac{\alpha}{n} L_{\rm att}+ \frac{\beta}{n} L_{\rm rep}-\frac{\mathbf{1} \mathbf{1}^T}{n}\Big)z\nonumber\\
&=\max_{|z|=1} \bigg[ 1-\frac{\alpha}{n}\sum_{(i,j)\in \mathcal{E}_{\rm att}, i<j} p_{ij} (z_i-z_j)^2\nonumber\\
&\ +\frac{\beta}{n}\sum_{(i,j)\in \mathcal{E}_{\rm rep}, i<j} p_{ij} (z_i-z_j)^2+\frac{1}{n}\big(\sum_{i=1}^n z_i \big)^2 \bigg] \nonumber\\
& \triangleq \max_{|z|=1} H_{\alpha,\beta, \mathcal{G}_{\rm att}, \mathcal{G}_{\rm rep}}(z),
\end{align}
where $z=(z_1 \dots z_n)^T\in \mathds{R}^n$. It is now clear that $H_{\alpha_1,\beta, \mathcal{G}_{\rm att}, \mathcal{G}_{\rm rep}}(z)\geq H_{\alpha_2,\beta, \mathcal{G}_{\rm att}, \mathcal{G}_{\rm rep}}(z)$ for any $0\leq \alpha_1\leq \alpha_2\leq 1$, and that $H_{\alpha,\beta_1, \mathcal{G}_{\rm att}, \mathcal{G}_{\rm rep}}(z)\geq H_{\alpha,\beta_2, \mathcal{G}_{\rm att}, \mathcal{G}_{\rm rep}}(z)$ for any $0\leq \beta_1\leq \beta_2$. This implies the desired result in light of (\ref{5}).
\hfill$\square$

\end{document}